\documentclass[aps,pre,floatfix]{revtex4}
\usepackage{epsf}
\usepackage{subfigure}
\usepackage{amsmath}
\usepackage{amssymb}
\usepackage{graphicx}
\usepackage{epstopdf}
\begin{document}

\title{\bf Fluctuation theorem for currents and nonlinear response
coefficients}

\author{David Andrieux and Pierre Gaspard}
\affiliation{Center for Nonlinear Phenomena and Complex Systems,\\
Universit\'e Libre de Bruxelles, Code Postal 231, Campus Plaine,
B-1050 Brussels, Belgium}

\begin{abstract}
We use a recently proved fluctuation theorem for the
currents to develop the response theory of nonequilibrium phenomena.
In this framework, expressions for the response coefficients of the
currents at arbitrary orders in the thermodynamic forces or
affinities are obtained in terms of the fluctuations of the
cumulative currents and remarkable relations are obtained
which are
the consequences of microreversibility beyond Onsager reciprocity
relations.
\end{abstract}

\maketitle

\section{Introduction}

Onsager's classic work of 1931 \cite{O31}
has
shown that the linear response coefficients relating the currents
to
the thermodynamic forces or affinities \cite{DD36} obey reciprocity
relations as a consequence
of the reversibility of the underlying
microscopic dynamics.
Another consequence of microreversibility are
the so-called fluctuation theorems,
which characterize the large
fluctuations of physical quantities in nonequilibrium systems.
They
have been derived in deterministic \cite{ECM93,ES94,GC95,G96} or
Markovian stochastic systems \cite{K98,LS99,M99,MN03} and concern
different quantities such as the entropy production \cite{GC95}, the
dissipated work \cite{C99,ZC03}, or the currents crossing the system
in a nonequilibrium situation \cite{AG04,AG05,AG06a}.  Such relations
are important because they are valid far from equilibrium.
Close
enough to equilibrium where the response of the system is linear in
the affinities, the Onsager reciprocity relations can be deduced from the
fluctuation theorem \cite{G96,LS99,AG04}.

On the other hand, it is
known that far-from-equilibrium systems may present nonlinear
responses
to nonequilibrium constraints.  The response is said to be
nonlinear
if the currents crossing the nonequilibrium system depend
nonlinearly on the affinities.
The coefficients characterizing such
nonlinear responses
can be obtained by expanding the currents in the
powers of the affinities.
The terms linear in the affinities are the
linear response coefficients obeying Onsager's reciprocity relations.
The terms which are quadratic, cubic, quartic, etc... in the
affinities are called the
nonlinear response coefficients.  We may
wonder if the nonlinear response coefficients would obey relations
beyond Onsager's ones as the conequence of the fundamental
microreversibility.

The purpose of the present paper is to show
that, indeed, the nonlinear response coefficients do obey
remarkable
relations which have their origin in microreversibility.
For this
purpose, we use a fluctuation theorem for the currents which was 
first proven for mechanically driven Markovian processes 
\cite{LS99,M99,MN03}, then in the more general framework of
Schnakenberg network theory \cite{S76} which includes reactive 
processes
\cite{AG04,AG05}, as well as in non-Markovian situations 
\cite{AG06c}.  
This fluctuation theorem directly concerns
the generating function of the different fluctuating currents
crossing a nonequilibrium system.  Consequently, the nonlinear
response coefficients can be directly obtained from the generating
function by successive differentiations, so that the symmetry of the
fluctuation theorem for the currents can be used in a straightforward
way.  The fluctuation theorem for the currents has been proved
elsewhere \cite{AG05} under the general conditions 
enunciated by 
Schnakenberg \cite{S76}
and we start from this important result to
obtain remarkable relations as the consequences of
microreversibility.

The plan of the paper is the following.  In
Section \ref{Summary}, we summarize the results about the fluctuation
theorem for the currents.  Section \ref{R123} is devoted to the
derivation of the consequences of the fluctuation theorem on the
response coefficients up to the cubic response coefficients with
comparisons with known results.  In Section \ref{SYS}, the nonlinear
response coefficients are systematically calculated and the
generalizations of Onsager relations are obtained at arbitrarily
large orders.  Conclusions are drawn in Section \ref{Conclusions}.


\section{Fluctuation theorem for the
currents}
\label{Summary}

The fluctuation theorem for the currents relates the probability of 
observing given
values for the cumulated currents to the probability of observing
negative values via the following exponential relation valid in the
long-time limit:
\begin{equation}
\frac{ \text{Prob}\left[\left\{ \frac{1}{t}\int_0^t dt' j_{\gamma}(t')
\in \left(\xi_{\gamma},\xi_{\gamma}+d\xi_{\gamma}\right)
\right\} \right] }{\text{Prob}\left[\left\{ \frac{1}{t}\int_0^t dt'
j_{\gamma}(t') \in
\left(-\xi_{\gamma},-\xi_{\gamma}+d\xi_{\gamma}\right) \right\}
\right] }
\simeq \exp{ \sum_{\gamma} A_{\gamma} \xi_{\gamma} t}
\qquad\qquad (t \rightarrow \infty )
\label{CFTrap}
\end{equation}
where $j_{\gamma}(t)$ denote the independent fluctuating currents and
$A_{\gamma}$ are the corresponding affinities (also called the
thermodynamic forces) driving the system out of equilibrium
\cite{DD36,S76}.

If we introduce the decay rate of the probability
that the cumulated currents take given values
\begin{equation}
H(\{ \xi_{\gamma} \} ) \equiv
\lim_{t \rightarrow \infty} - \frac{1}{t} \ln \text{Prob}\left[
\left\{  \frac{1}{t} \int_{0}^{t} dt' j_{\gamma} (t')
\in \left( \xi_{\gamma},\xi_{\gamma}+d\xi_{\gamma}\right)\right\}
\right]
\label{decay.rate}
\end{equation}
the fluctuation theorem can be written as
\begin{equation}
H(\{- \xi_{\gamma} \} ) - H(\{ \xi_{\gamma} \} )  =
\sum_{\gamma} A_{\gamma} \xi_{\gamma}
\label{FT.new.tl}
\end{equation}

The Legendre transform of the decay rate (\ref{decay.rate})
\begin{equation}
Q(\{\lambda_{\gamma}\}) ={\rm Max}_{\{\xi_{\gamma}\}} \left[
H(\{ \xi_{\gamma} \} ) +  \sum_{\gamma}
\lambda_{\gamma}\xi_{\gamma}\right]
\end{equation}

is the generating function of the currents defined by
\begin{equation}
Q(\{\lambda_{\gamma} \}; \{ A_{\gamma} \}) \equiv
\lim_{t\to\infty} -\frac{1}{t} \ln \left \langle {\rm e}^{- \sum_{\gamma}
\lambda_{\gamma} G_{\gamma}(t)}\right \rangle
\label{gen.fn}
\end{equation}
in terms of the cumulative currents also called the
Helfand moments \cite{H60}:
\begin{equation}
G_{\gamma}(t) \equiv \int_0^{t}  j_{\gamma} (t') \; dt'
\label{Helfand}
\end{equation}

The fluctuation theorem for the currents
(\ref{CFTrap}) is now expressed as
\begin{equation}
Q(\{\lambda_{\gamma} \}; \{ A_{\gamma} \}) =
Q(\{ A_{\gamma} -
\lambda_{\gamma} \}; \{ A_{\gamma} \})
\label{CFT}
\end{equation}
in terms of the generating function.  This
relation has been derived in the context of stochastic processes
\cite{AG05}. In this description the system is described by a
probability distribution over the possible states and which is ruled
by a master equation with several transition rates. The macroscopic
affinities $A_{\gamma}$ are then identified using Schnakenberg's
network theory \cite{S76}.

In the nonequilibrium steady state, the
mean value of the current
$j_{\alpha} (t)$ is given by
\begin{equation}
J_{\alpha} \equiv  \frac{\partial Q}{\partial
\lambda_{\alpha}}\Big\vert_{\{ \lambda_{\gamma}=0 \}}
= \lim_{t\to\infty} \frac{1}{t}\int_0^t \langle j_{\alpha}(t')\rangle \; dt'
= \lim_{t\to\infty} \frac{1}{t}\langle G_{\alpha}(t) \rangle
\label{q.macro.flux}
\end{equation}


\section{Consequences for nonlinear response}
\label{R123}

In this section, we prove that the fluctuation
theorem (\ref{CFT}) for the macroscopic currents
(\ref{q.macro.flux}) have important consequences
not only on the linear response coefficients but also
at the level of the nonlinear response.
In general, the macroscopic currents can be expanded as power series
of the macroscopic affinities:
\begin{equation}
J_{\alpha} = \sum_{\beta} L_{\alpha\beta} A_{\beta} + \frac{1}{2}
\sum_{\beta,\gamma} M_{\alpha\beta\gamma} A_{\beta}
A_{\gamma}
+ \frac{1}{6} \sum_{\beta,\gamma,\delta} N_{\alpha\beta\gamma\delta}
A_{\beta} A_{\gamma} A_{\delta} + \cdots
\label{macro.fluxes-affinities}
\end{equation}
The linear response of the currents $J_{\alpha}$ with respect to a small
perturbation in the affinities $A_{\beta}$
is characterized by the Onsager coefficients $L_{\alpha\beta}$, and
the nonlinear response by the higher-order coefficients
$M_{\alpha\beta\gamma}$, $N_{\alpha\beta\gamma\delta}$,...

\subsection{Onsager reciprocity relations}

The Onsager coefficients are defined close to the equilibrium in
terms of the generating function (\ref{CFT}) by
\begin{equation}
L_{\alpha\beta} \equiv \frac{\partial J_{\alpha}}{\partial
A_{\beta}}\Big\vert_{\pmb{A}=0}
= \frac{\partial^2 Q}{\partial \lambda_{\alpha}\partial A_{\beta}}(0;0)
\label{L}
\end{equation}
If we differentiate the expression (\ref{CFT}) of the
fluctuation theorem with respect to $\lambda_{\alpha}$ and $
A_{\beta}$
we find that
\begin{equation}
\frac{\partial^2 Q}{\partial \lambda_{\alpha}\partial
A_{\beta}}(\pmb{\lambda};\pmb{A}) =
- \frac{\partial^2 Q}{\partial \lambda_{\alpha}\partial
\lambda_{\beta}}(\pmb{A}-\pmb{\lambda};\pmb{A})
- \frac{\partial^2 Q}{\partial \lambda_{\alpha}\partial
A_{\beta}}(\pmb{A}-\pmb{\lambda};\pmb{A})
\label{diff2}
\end{equation}
Setting $\pmb{\lambda}=0$ and $\pmb{A}=0$, we
obtain the relation
\begin{equation}
2 \frac{\partial^2 Q}{\partial \lambda_{\alpha}\partial
A_{\beta}}(0;0) =
- \frac{\partial^2 Q}{\partial \lambda_{\alpha}\partial \lambda_{\beta}}(0;0)
    \label{diff2.0}
\end{equation}
or
\begin{equation}
L_{\alpha\beta}= - \frac{1}{2}\frac{\partial^2 Q}{\partial
\lambda_{\alpha}\partial \lambda_{\beta}}(0;0)
\label{L.diff2}
\end{equation}
as already shown in reference \cite{G96,LS99}.
Hence the Onsager reciprocity relations
\begin{equation}
L_{\alpha\beta}= L_{\beta\alpha}
\label{ORR}
\end{equation}
We notice that no further relation is obtained by differentiating the
fluctuation relation
(\ref{CFT}) twice with respect to either
the parameters $\pmb{\lambda}$ or the affinities $\pmb{A}$.

\subsection{Green-Kubo and Einstein-Helfand formulas}

By using Eq. (\ref{L.diff2}),
we obtain the Onsager coefficients as
\begin{equation}
L_{\alpha\beta}= \frac{1}{2}\int_{-\infty}^{+\infty} \langle \left[
j_{\alpha}(t)-\langle j_{\alpha}\rangle\right]
\left[ j_{\beta}(0)-\langle j_{\beta}\rangle\right]\rangle_{\rm eq} \; dt
=  \lim_{t\to\infty} \frac{1}{2t} \, \langle \Delta
G_{\alpha}(t)\Delta G_{\beta}(t)\rangle_{\rm eq}
\label{YZ}
\end{equation}
in terms of the time correlation
functions of the instantaneous currents or the corresponding Helfand moments:
\begin{equation}
\Delta G_{\alpha}(t) \equiv G_{\alpha}(t) - \langle G_{\alpha}(t)\rangle
\end{equation}
Here, the statistical average is carried out with respect to the
state of thermodynamic equilibrium.
In Eq. (\ref{YZ}), the formulas giving the coefficients
in terms of the time correlation functions are known
as the Green-Kubo formulas \cite{G52,K57}
(or the Yamamoto-Zwanzig formulas in the context of chemical
reactions \cite{Y60,Z65}). The other formulas giving
the coefficients in terms of the Helfand moments or cumulative currents
are known as the Einstein-Helfand formulas \cite{H60,E05}.

\subsection{Relations for the second-order response coefficients}

The second-order response coefficients are defined as
\begin{equation}
M_{\alpha\beta\gamma} \equiv
\frac{\partial^3 Q}{\partial \lambda_{\alpha}\partial
A_{\beta}\partial A_{\gamma}}(0;0)
\label{M3}
\end{equation}
in terms of one derivative with respect to the parameter
$\lambda_{\alpha}$ generating the current $J_{\alpha}$
and two derivatives
with respect to the affinities $A_{\beta}$ and $A_{\gamma}$.

Our purpose is to relate these nonlinear response coefficients
to quantities with a reduced number of derivatives
with respect to the affinities, thus characterizing the fluctuations
instead of the response.

Such relations are obtained by continuing the procedure
started to get the Onsager reciprocity relations by further differentiating
the generating function.
If we differentiate the identity (\ref{diff2}) with respect to
$A_{\gamma}$ and set $\pmb{\lambda}=0$ and $\pmb{A}=0$,
we obtain the second-order response coefficients as
\begin{equation}
M_{\alpha\beta\gamma}= - \frac{1}{2}\frac{\partial^3 Q}{\partial
\lambda_{\alpha}\partial \lambda_{\beta}\partial\lambda_{\gamma}}(0;0)
- \frac{1}{2}\frac{\partial^3 Q}{\partial \lambda_{\alpha}\partial
\lambda_{\beta}\partial A_{\gamma}}(0;0)
- \frac{1}{2}\frac{\partial^3 Q}{\partial \lambda_{\alpha}\partial
\lambda_{\gamma}\partial A_{\beta}}(0;0)
\label{M.diff3}
\end{equation}

Using the symmetry (\ref{CFT}) of the current
fluctuation theorem at equilibrium,
we see that the generating function $Q(\pmb{\lambda}; 0 ) =
Q(- \pmb{\lambda}; 0 ) $
is an even function of ${\lambda_\gamma}$ at equilibrium.
The first term of the right-hand side of equation (\ref{M.diff3}) is
a third derivative with respect to the parameters ${\lambda_\gamma}$
calculated at equilibrium, hence it must vanish identically.

The other terms in Eq. (\ref{M.diff3}) are related to
the spectral function of the nonequilibrium current fluctuations
defined by
\begin{equation}
\Sigma_{\alpha\beta}(\omega) \equiv \int_{-\infty}^{+\infty}
{\rm e}^{i\omega t} \, \langle \left[ j_{\alpha}(t)-\langle
j_{\alpha}\rangle\right]
\left[ j_{\beta}(0)-\langle j_{\beta}\rangle\right]\rangle \; dt
\label{spectrum}
\end{equation}
where the statistical average is here taken with respect to the
nonequilibrium steady state.
Here, we introduce the quantities
\begin{eqnarray}
R_{\alpha\beta, \gamma} &\equiv& - \frac{\partial^3 Q}{\partial
\lambda_{\alpha}\partial
\lambda_{\beta}\partial A_{\gamma}}(0;0)  \nonumber \\ &=&
\frac{\partial}{\partial A_{\gamma}}\int_{-\infty}^{+\infty}
\langle \left[ j_{\alpha}(t)-\langle j_{\alpha}\rangle\right]
\left[ j_{\beta}(0)-\langle j_{\beta}\rangle\right]\rangle
\; dt\Big\vert_{\pmb{A}=0}
\nonumber \\ &=&
\frac{\partial}{\partial{A}_{\gamma}}\lim_{t\to\infty} \frac{1}{t}
\langle \Delta G_{\alpha}(t)\Delta
G_{\beta}(t)\rangle\Big\vert_{\pmb{A}=0}
\nonumber \\ &=&
\frac{\partial}{\partial A_{\gamma}}
\Sigma_{\alpha\beta}(\omega=0)\Big\vert_{\pmb{A}=0}
\label{R}
\end{eqnarray}
which characterize the sensitivity of the current fluctuations out of
equilibrium.
Equation (\ref{R}) shows that the sensitivity coefficients are given
in terms of
the derivative with respect to the affinities of the spectral
function or, equivalently,
of the diffusivities of the nonequilibrium currents
defined by
\begin{equation}
D_{\alpha\beta} \equiv
\lim_{t\to\infty} \frac{1}{2t} \langle \Delta G_{\alpha}(t)\Delta
G_{\beta}(t)\rangle
= - \frac{1}{2} \frac{\partial^2 Q}{\partial
\lambda_{\alpha}\partial \lambda_{\beta}}\Big\vert_{\pmb{\lambda}=0}
\label{D}
\end{equation}

According to the fluctuation theorem and Eq. (\ref{M.diff3}),
we find that the second-order response coefficients are given in
terms of the sensitivity coefficients by
\begin{equation}
M_{\alpha\beta\gamma} = \frac{1}{2} \left(R_{\alpha\beta,\gamma} +
R_{\alpha\gamma,\beta}\right)
\label{RR3}
\end{equation}
For the case $\beta = \gamma$ we find
\begin{equation}
M_{\alpha\beta\beta} = R_{\alpha\beta,\beta}
\label{RR3b}
\end{equation}
In particular the response coefficients $M_{\alpha\beta\gamma}$
present the expected symmetry $M_{\alpha\beta\gamma} =
M_{\alpha\gamma\beta}$. The second-order coefficients are thus
related to the diffusivities by
\begin{equation}
M_{\alpha\beta\gamma} =  \left[\frac{\partial}{\partial A_{\gamma}}
D_{\alpha\beta} +
\frac{\partial}{\partial A_{\beta}} D_{\alpha\gamma}\right]_{\pmb{A}=0}
\label{M.D}
\end{equation}
Thanks to the fluctuation theorem for the currents, we can therefore relate
the second-order nonlinear response coefficients to quantities characterizing
the nonequilibrium fluctuations such as the spectral functions or the
diffusivities
of the currents in the nonequilibrium steady state.  We notice that
the number of derivatives
with respect to the affinities has indeed been reduced.

Similar expressions can be found for even higher-order relations
where the odd derivatives with respect to the $\lambda$'s
automatically vanish at equilibrium.

\subsection{Relations for the third-order response coefficients}

A similar reasoning can be carried out for the third-order response
coefficients defined by
\begin{equation}
N_{\alpha\beta\gamma\delta} \equiv
\frac{\partial^4 Q}{\partial \lambda_{\alpha}\partial
A_{\beta}\partial A_{\gamma}\partial A_{\delta}}(0;0)
\label{N4}
\end{equation}

Differentiating the identity (\ref{diff2}) twice with
respect to the
affinities $A_{\gamma}$ and $A_{\delta}$ shows that
\begin{equation}
N_{\alpha\beta\gamma\delta}= - \frac{1}{2}\frac{\partial^4
Q}{\partial \lambda_{\alpha}\partial
\lambda_{\beta}\partial\lambda_{\gamma}\partial\lambda_{\delta}}(0;0)
- \frac{1}{2}\left( S_{\alpha\beta\gamma,\delta} +
S_{\alpha\beta\delta,\gamma} + S_{\alpha\gamma\delta,\beta}\right)
+ \frac{1}{2}\left( T_{\alpha\beta,\gamma\delta} +
T_{\alpha\gamma,\beta\delta} +
T_{\alpha\delta,\beta\gamma} \right)
\label{N.diff4}
\end{equation}
with
\begin{equation}
S_{\alpha\beta\gamma,\delta} \equiv
\frac{\partial^4 Q}{\partial \lambda_{\alpha}\partial
\lambda_{\beta}\partial\lambda_{\gamma}\partial A_{\delta}}(0;0)
\label{S}
\end{equation}
and
\begin{equation}
T_{\alpha\beta,\gamma\delta} \equiv
- \frac{\partial^4 Q}{\partial \lambda_{\alpha}\partial
\lambda_{\beta}\partial A_{\gamma}\partial
A_{\delta}}(0;0)
\label{T}
\end{equation}
Differentiating Eq. (\ref{diff2}) twice with respect to the
parameters $\lambda_{\gamma}$ and $\lambda_{\delta}$
shows that
\begin{equation}
S_{\alpha\beta\gamma,\delta} = - \frac{1}{2}
\frac{\partial^4 Q}{\partial \lambda_{\alpha}\partial
\lambda_{\beta}\partial\lambda_{\gamma}\partial\lambda_{\delta}}(0;0)
\label{Ssym}
\end{equation}
which proves the total symmetry of this tensor.
Accordingly, the third-order response coefficients are given by
\begin{equation}
N_{\alpha\beta\gamma\delta}= - \frac{1}{2} S_{\alpha\beta\gamma,\delta}
+ \frac{1}{2}\left( T_{\alpha\beta,\gamma\delta} +
T_{\alpha\gamma,\beta\delta} +
T_{\alpha\delta,\beta\gamma} \right)
\label{N.diff4.bis}
\end{equation}

We thus obtain the reciprocity relations that the fourth-order tensor
\begin{equation}
2 N_{\alpha \beta \gamma \delta} - T_{\alpha \beta,\gamma \delta}
- T_{\alpha \gamma,\beta \delta} - T_{\alpha \delta,\beta \gamma}
\label{RR4}
\end{equation}
must be totally symmetric.

The tensor (\ref{S}) can be expressed as
\begin{equation}
S_{\alpha\beta\gamma,\delta}  =
\frac{\partial}{\partial A_{\delta}}\lim_{t\to\infty} \frac{1}{t}
\langle \Delta G_{\alpha}(t)\Delta G_{\beta}(t)\Delta G_{\gamma}(t)
\rangle\Big\vert_{\pmb{A}=0}
\label{S.G}
\end{equation}
which do not vanish in general as for the even-order cases.
Equation (\ref{S.G}) shows that the tensor (\ref{S})
characterizes the sensitivity of the third-order moments
of the cumulative currents with respect to the nonequilibrium constraints.
Moreover,  the expression (\ref{Ssym}) shows
that this tensor can also be calculated at  equilibrium as
\begin{eqnarray}
S_{\alpha\beta\gamma,\delta} = \lim_{t\to\infty} &&\frac{1}{2t}
\Big[\langle \Delta G_{\alpha}(t)\Delta G_{\beta}(t)\Delta
G_{\gamma}(t)\Delta G_{\delta}(t)\rangle
\nonumber\\ && \qquad -\langle \Delta G_{\alpha}(t)\Delta G_{\beta}(t)\rangle
\langle\Delta G_{\gamma}(t)\Delta G_{\delta}(t)\rangle
\nonumber\\ && \qquad -\langle \Delta G_{\alpha}(t)\Delta G_{\gamma}(t)\rangle
\langle\Delta G_{\beta}(t)\Delta G_{\delta}(t)\rangle
\nonumber\\ && \qquad -\langle \Delta G_{\alpha}(t)\Delta G_{\delta}(t)\rangle
\langle\Delta G_{\beta}(t)\Delta G_{\gamma}(t)\rangle\Big]_{\rm eq}
\label{Ssym.moments}
\end{eqnarray}
where it characterizes the fluctuations.  The equality between Eqs.
(\ref{S.G}) and
(\ref{Ssym.moments}) is another remarkable consequence of the
fluctuation theorem.

On the other hand, the tensor (\ref{T}) is given by
\begin{eqnarray}
T_{\alpha\beta,\gamma\delta}  &=&
\frac{\partial}{\partial
A_{\gamma}}\frac{\partial}{\partial A_{\delta}}
\int_{-\infty}^{+\infty}
\langle \left[ j_{\alpha}(t)-\langle j_{\alpha}\rangle\right]
\left[ j_{\beta}(0)-\langle j_{\beta}\rangle\right]\rangle
\; dt\Big\vert_{\pmb{A}=0}\nonumber \\
&=&
\frac{\partial}{\partial
A_{\gamma}}\frac{\partial}{\partial A_{\delta}}
\lim_{t\to\infty}
\frac{1}{t} \langle \Delta G_{\alpha}(t)\Delta G_{\beta}(t)\rangle
\Big\vert_{\pmb{A}=0} \nonumber \\
&=&
\frac{\partial}{\partial
A_{\gamma}}\frac{\partial}{\partial A_{\delta}}
\Sigma_{\alpha\beta}(\omega=0)\Big\vert_{\pmb{A}=0}
\label{T.correl}
\end{eqnarray}
Accordingly, the tensor (\ref{T}) also characterizes the sensitivity
of the nonequilibrium fluctuations but now in terms of the second derivatives
of the power spectrum with respect to the affinities.
We notice the similarity with Eq. (\ref{R}).  Again, the number of
derivatives with respect to
the affinities has been reduced compared to the definition (\ref{N4}) of
the third-order response coefficients and this thanks to the
fluctuation theorem for the
currents.

The expansion can be carried out to higher orders as done in the next section.


\section{Relations at arbitrary orders}
\label{SYS}

In a macroscopic description, we consider general affinities
$A_{\alpha} $ conjugated to currents $J_{\alpha}$.
The mean value of the currents can be developed as
\begin{equation}
J_{\alpha} = \sum_{n=1}^{\infty} \frac{1}{n!} \sum_{\beta,\cdots,\mu}
C^{(n)}_{\alpha\beta\cdots\mu} \underbrace{A_{\beta} \dots A_{\mu}}_{n}
\label{flux}
\end{equation}
with the
coefficients
\begin{equation}
C^{(n)}_{\alpha\beta\cdots\mu} \equiv
\frac{\partial^n J_{\alpha}}{\partial A_{\beta}\cdots \partial
A_{\mu}}\Big\vert_{\pmb{A}=0} = \frac{\partial^{n+1} Q}{\partial
\lambda_{\alpha}\partial A_{\beta}\cdots \partial
A_{\mu}}\Big\vert_{\pmb{\lambda}=0,\pmb{A}=0}
\label{coeff}
\end{equation}
contain $n+1$ indices. The expansion in powers of the
affinities gives Onsager's
linear response coefficients $C^{(1)}_{\alpha\beta}$
as well as higher-order coefficients $C^{(n)}$ characterizing the nonlinear
response of the system with respect to the nonequilibrium constraints
$\{ A_{\epsilon} \}$.

We now want to use the fluctuation theorem for the currents
(\ref{CFT}) to obtain expressions for the response coefficients. To
do so we will consider expressions for the derivatives of $Q$ at
arbitrary orders. This will provide us with several non trivial
relationships and we will have to combine them to obtain a simple
form for the response coefficients.

Using the fluctuation theorem (\ref{CFT}), the derivatives of $Q$ are given by
\begin{equation}
Q_{\alpha\cdots\eta,\rho\cdots\sigma}^{(k,n)} =
(-1)^k \ \sum_{p=0}^{n-k} \
Q_{\alpha\cdots\eta\{\rho\cdots\sigma\}_p}^{(k+p,n)}
\label{DQ}
\end{equation}
where the derivatives are calculated at $\pmb{\lambda} = \pmb{A} = 0$.
The notation $Q_{\alpha\cdots\eta,\rho\cdots\sigma}^{(k,n)}$ means
that we have taken $n$ derivatives with $k$ of them corresponding to
$\lambda_{\alpha}$,...,$\lambda_{\eta}$ and $n-k$ corresponding to
$A_{\rho}$,...,$A_{\sigma}$:
\begin{equation}
Q_{\alpha\cdots\eta,\rho\cdots\sigma}^{(k,n)} \equiv
\frac{\partial^n Q}{\underbrace{\partial \lambda_{\alpha} \cdots
\partial \lambda_{\eta}}_k \underbrace{\partial A_{\rho} \cdots
\partial A_{\sigma}}_{n-k}}(0;0)
\end{equation}
These derivatives correspond to the response of the $k^{\rm th}$
cumulant with respect to the macroscopic affinities:
\begin{equation}
Q_{\alpha\cdots\eta,\rho\cdots\sigma}^{(k,n)} \equiv (-1)^{k+1}
\frac{\partial^{n-k} }{\partial A_{\rho} \cdots \partial A_{\sigma}} 
\lim_{t\rightarrow\infty} \frac{1}{t}
\langle\langle G_{\alpha}\cdots G_{\eta}  \rangle\rangle
\Big\vert_{\pmb{A}=0}
\end{equation}
where $\langle\langle \cdot \rangle\rangle$ denotes the cumulant.
The notation $\{ \cdot \}_p$ means the symmetrized ensemble
  with the
derivatives taken after the term $p$.
For example, $\{\alpha\beta\gamma\delta\}_1 \equiv
\alpha,\beta\gamma\delta + \beta,\alpha\gamma\delta +
\gamma,\alpha\beta\delta + \delta,\alpha\beta\gamma$.
In the same way, $\{\alpha\beta\gamma\delta\}_2 \equiv
\alpha\beta,\gamma\delta + \alpha\gamma,\beta\delta +
\alpha\delta,\gamma\beta + \beta\gamma,\alpha\delta+
\beta\delta,\alpha\gamma+\gamma\delta,\alpha\beta$.
There are $m!/[p!(m-p)!]$ numbers of terms if there are $m$ terms in
the ensemble.
The derivative (\ref{DQ}) is thus expressed as the sum of $2^{n-k}$
terms. The sign $(-1)^k$ comes from the derivatives with respect to
$\lambda$ while the structure of the sum comes from the derivatives
with respect to $A$.
Indeed, each such derivative generates two terms, one with a
derivative with respect to $A$ and the other with respect to
$\lambda$ as can be seen from Eq.~(\ref{CFT}).
Noting that the term $p=0$ in Eq. (\ref{DQ}) is the same as the
left-hand side of the equation, we have
\begin{equation}
\left[ 1+(-1)^{k+1}\right] \ Q_{\alpha\cdots\eta,\rho\cdots\sigma}^{(k,n)} =
(-1)^k \ \sum_{p=1}^{n-k} \
Q_{\alpha\cdots\eta\{\rho\cdots\sigma\}_p}^{(k+p,n)}
\end{equation}
so that
\begin{eqnarray}
0 &=& \ \sum_{p=1}^{n-k} \
Q_{\alpha\cdots\eta\{\rho\cdots\sigma\}_p}^{(k+p,n)} \quad k \;
\mbox{even} \label{simpl1}\\
2 \ Q_{\alpha\cdots\eta,\rho\cdots\sigma}^{(k,n)} &=& - \
\sum_{p=1}^{n-k} \
Q_{\alpha\cdots\eta\{\rho\cdots\sigma\}_p}^{(k+p,n)} \quad k \;
\mbox{odd}
\label{simpl}
\end{eqnarray}
As explained above each derivative $Q^{(k,n)}$ can be expressed as
derivatives with respect to $n-k$ affinities of a cumulant of order
$k$ at equilibrium. We thus have a number of non trivial relations
between different moments and their derivatives with respect to the
affinities calculated at equilibrium.

In particular, the response coefficients (\ref{coeff}) of order $n-1$
are given
  by
  \begin{equation}
C_{\alpha\beta\cdots\sigma}^{(n-1)} = Q_{\alpha,\beta\cdots\sigma}^{(1,n)} =
- \frac{1}{2} \ \sum_{p=1}^{n-1} \ Q_{\alpha\{\beta\cdots\sigma\}_p}^{(1+p,n)}
\label{RC2}
\end{equation}
where we used Eq. (\ref{simpl}) with $k=1$ which is odd.
The response coefficients are thus expressed as a sum of $2^{n-1}-1$
terms with $n-1$ different tensors.

The relations (\ref{DQ}) can be used to simplify the expressions of
the response coefficients. For example, the relations
(\ref{simpl1})-(\ref{simpl}) with $k=n-1$ give
\begin{eqnarray}
  Q_{\alpha\cdots\sigma,}^{(n,n)} &=& 0 \quad n \; \mbox{odd}
\label{n-1}\\
2 \ Q_{\alpha\cdots\eta,\sigma}^{(n-1,n)} &=& - \
Q_{\alpha\cdots\sigma,}^{(n,n)} \quad n \; \mbox{even}
\label{n-1even}
\end{eqnarray}
so that the derivatives with respect to $\lambda$ vanishes if $n$ is
odd
and we obtain the total symmetry of the tensor
$Q_{\alpha\cdots\eta,\sigma}^{(n-1,1)}$
if $n$ is even:
\begin{equation}
\frac{\partial}{\partial A_{\sigma}} \lim_{t\rightarrow\infty} 
\frac{1}{t} \langle\langle G_{\alpha}\cdots
G_{\eta}
\rangle\rangle \Big\vert_{\pmb{A}=0} = \frac{1}{2} 
\lim_{t\rightarrow\infty} \frac{1}{t}
\langle\langle G_{\alpha}\cdots G_{\eta} G_{\sigma}  \rangle\rangle_{\rm eq}
\end{equation}
which is totally symmetric. These are non-trivial consequences 
derived from the fluctuation theorem.
If $k=n$ we do not get any new informations.
If $k=0$ we find a constraint on the sum of the response coefficients
but it can be recovered from their expressions (\ref{RC2}). In fact
all relations (\ref{simpl1}) with $k$ even can be recovered from
relations (\ref{simpl}) for $k$ odd. We now have to use them to
simplify the expression of the response coefficients.

To do this, we will thus choose to use relations (\ref{simpl}) to
eliminate all terms of the form $Q^{(k,n)}$ where $k$ is odd.
However, the terms in the right hand side of Eq. (\ref{RC2}) are all 
totally symmetrized wrt
the $n-1$ indexes $\{ \beta\cdots\sigma\}$ while relations
(\ref{simpl}) are not. The first step is thus to symmetrize relations
(\ref{simpl}) to get
\begin{eqnarray}
Q_{\alpha\{\beta\cdots\sigma\}_{k-1}}^{(k,n)} = - \frac{1}{2} \ 
\sum_{l=k}^{n-1}
\ \left(\begin{array}{c} l \\ k-1 \end{array}\right) \
Q_{\alpha\{\beta\cdots\sigma\}_{l}}^{(l+1,n)} \quad \mbox{for $k$ odd}
\label{Qsym}
\end{eqnarray}
where
\begin{eqnarray}
\left(\begin{array}{c} l \\ k-1 \end{array}\right) \equiv
\frac{l!}{(k-1)!(l-k+1)!}
\end{eqnarray}
These coefficients are obtained by symmetrizing the relations (\ref{simpl}).
As all relations become thus totally symmetric, there are
$\left(\begin{array}{c} l \\ k-1 \end{array}\right)$ identically
terms $Q_{\alpha\{\beta\cdots\sigma\}_{l}}^{(l+1,n)}$ arising by this
procedure.
These coefficients are given by the numbers of terms in
$Q_{\alpha\{\beta\cdots\sigma\}_{k-1}}^{(k,n)}$ times the number of
terms in $Q_{\alpha\cdots\eta\{\rho\cdots\sigma\}_{l-k+1}}^{(l+1,n)}$
divided by the number of terms in
$Q_{\alpha\{\beta\cdots\sigma\}_{l}}^{(l+k,n)}$. Unfortunately
expression (\ref{Qsym}) is expressed in terms of $Q^{(k,n)}$ with $k$
odd and even. We will thus have to use this relation recursively to
eliminate all terms with $k$ odd within itself.

One have then to eliminate successively the terms
$Q_{\alpha\{\beta\cdots\sigma\}_{2}}^{(3,n)}$,
$Q_{\alpha\{\beta\cdots\sigma\}_{4}}^{(5,n)}$,...,$Q_{\alpha\{\beta\cdots\sigma\}_{p-1}}^{(p,n)}$. 

When this is done, one can express the response coefficients in the
form
\begin{equation}
C_{\alpha\beta\cdots\sigma}^{(n-1)} =
\gamma_1 \ Q_{\alpha\{\beta\cdots\sigma\}_1}^{(2,n)} + \gamma_3 \
Q_{\alpha\{\beta\cdots\sigma\}_3}^{(4,n)}+\cdots
+ \gamma_{n-2} \ Q_{\alpha\{\beta\cdots\sigma\}_{n-2}}^{(n-1,n)}
\qquad (n \; \mbox{odd})
\label{RCodd}
\end{equation}
and \begin{equation}
C_{\alpha\beta\cdots\sigma}^{(n-1)} =
\gamma_1 \ Q_{\alpha\{\beta\cdots\sigma\}_1}^{(2,n)} + \gamma_3 \
Q_{\alpha\{\beta\cdots\sigma\}_3}^{(4,n)}+\cdots
+ \gamma_{n-1} \
Q_{\alpha\beta\cdots\sigma}^{(n,n)}  \qquad (n \; \mbox{even})
\label{RCeven}
\end{equation}
where we used that $Q^{(n,n)} = 0$ for $n$ odd. We find here the
important property that the coefficients $\left(\begin{array}{c} l \\
k-1 \end{array}\right)$
are independent of $n$ which implies that
the coefficients $\gamma_i$ do not depend on $n$.
The odd response coefficients are thus expressed in terms of
$(n-1)/2$ different tensors which is far better than the $n-1$
tensor needed in expression (\ref{RC2}). The even response
coefficients are thus expressed in terms of $n/2$ different tensors.
By construction, they are symmetric for the permutations of the $n-1$
indexes $\{\beta\cdots\sigma\}$ as it should.

In particular, the tensor
\begin{equation}
C_{\alpha\beta\cdots\sigma}^{(n-1)} - \ \left[ \gamma_1 \
Q_{\alpha\{\beta\cdots\sigma\}_1}^{(2,n)} + \gamma_3 \
Q_{\alpha\{\beta\cdots\sigma\}_3}^{(4,n)} + \cdots +
  \gamma_{n-3} \
Q_{\alpha\{\beta\cdots\sigma\}_{n-3}}^{(n-2,n)}  \right]
=  \gamma_{n-1} \ Q_{\alpha\beta\cdots\sigma}^{(n,n)}
\end{equation}
with $n$ even is totally symmetric.

We now want to calculate the coefficients $\gamma_i$, with $i$ odd,
associated with the terms
$Q_{\alpha\{\beta\cdots\sigma\}_{i}}^{(i+1,n)}$. The first term 
$\gamma_1$ takes the value $-1/2$.
The next ones are given by the successive elimination of the terms
$Q_{\alpha\{\beta\cdots\sigma\}_{2}}^{(3,n)}$,
$Q_{\alpha\{\beta\cdots\sigma\}_{4}}^{(5,n)}$,...,
$Q_{\alpha\{\beta\cdots\sigma\}_{p-1}}^{(p,n)}$.
Each successive elimination will change the coefficients in front of 
the $Q_{\alpha\{\beta\cdots\sigma\}_{l}}^{(l+1,n)}$. We thus 
introduce numbers $\chi_k^l$ which denote the coefficients pondering 
the terms $Q_{\alpha\{\beta\cdots\sigma\}_{l}}^{(l+1,n)}$ at the 
$k^{\rm th}$ successive elimination.
 From Eq. (\ref{RC2}) we set $\chi_0^l = -1/2 \  \forall l$. The first 
step is thus to eliminate the term 
$Q_{\alpha\{\beta\cdots\sigma\}_{2}}^{(3,n)}$ so that
\begin{equation}
\chi_1^l  = \begin{cases} \chi_0^l - \frac{1}{2} 
\left(\begin{array}{c} l \\ 2 \end{array}\right) \chi_0^2 & \text{if 
$l \geq 3$} \\
                 0 & \text{$l = 2$} \\
                 \chi_0^l & \text{if $l=1$} \\
                 \end{cases}
\end{equation}
according to Eq. (\ref{Qsym}).
We can continue and eliminate the term 
$Q_{\alpha\{\beta\cdots\sigma\}_{4}}^{(5,n)}$ to get
\begin{equation}
\chi_2^l  = \begin{cases} \chi_1^l - \frac{1}{2} 
\left(\begin{array}{c} l \\ 4 \end{array}\right) \chi_1^4 & \text{if 
$l \geq 5$} \\
                 0 & \text{if $l = 4$} \\
                 \chi_1^l & \text{otherwise} \\
                 \end{cases}
\end{equation}
and after $k$ steps we have
\begin{equation}
\chi_k^l  = \begin{cases} \chi_{k-1}^l - \frac{1}{2} 
\left(\begin{array}{c} l \\ 2k \end{array}\right) \chi_{k-1}^{2k} & 
\text{if $l \geq 2k+1$} \\
                 0 & \text{$l = 2k$} \\
                 \chi_{k-1}^l & \text{otherwise} \\
                 \end{cases}
\end{equation}
These numbers are independent of $n$ as it should. The coefficients 
$\gamma_i$ in Eq. (\ref{RCodd})-(\ref{RCeven}) are then given by 
$\chi_k^i$ once they remain invariant that is when $k \geq (i-1)/2$. 
This construction can be summarized in the form :
\begin{equation}
\gamma_i = \frac{1}{4} \ \sum_{p=3}^{i,2} \ \xi_p \;
\left(\begin{array}{c} i \\ p-1 \end{array}\right) - \frac{1}{2}
\label{gamma}
\end{equation}
where the notation $\sum_{p=a}^{b,2}$ means that we sum from $p=a$ to
$b$ by step of $2$. We also absorbed in the expression of $\gamma_i$
the factor $-\frac{1}{2}$ in front of expression (\ref{RC2}). The
constant $-\frac{1}{2}$ is the contribution from the original terms
$Q^{(i+1,n)}$ in Eq. (\ref{RC2}).
The coefficients $\xi_p$ give the number of terms
$Q_{\alpha\{\beta\cdots\sigma\}_{p-1}}^{(p,n)}$ coming from the
elimination of the previous odd terms. They are given by
\begin{eqnarray}
\xi_p = 1 - \frac{1}{2} \ \sum_{l=3}^{p-2,2} \
\left(\begin{array}{c} p-1 \\ l-1 \end{array}\right) \;
\xi_l
\end{eqnarray}
We then find
\begin{equation}
\xi_3 = 1, \; \xi_5 = -2, \; \xi_7=\frac{17}{2}, \; \xi_9=-62, \;\ldots
\end{equation}
Injecting those numbers in relations (\ref{gamma}) we find
\begin{equation}
\gamma_1 = -\frac{1}{2} , \; \gamma_3  = \frac{1}{4}, \; \gamma_5  =
-\frac{1}{2},  \;
\gamma_7 = \frac{17}{8}, \;
\gamma_9 = -\frac{31}{2}, \ldots
\label{gammanum}
\end{equation}
A shorter relation in order to obtain the $\gamma_i$
, $i > 1$, is given by
\begin{eqnarray}
\gamma_i = \frac{1}{4} - \frac{1}{2} \ \sum_{l=3}^{i-2,2} \
\left(\begin{array}{c} i-1 \\ l-1 \end{array}\right) \;
\gamma_l
\end{eqnarray}

We have thus obtained expressions for the response coefficients of order $n$
in terms of microscopic correlation functions and their response to
the affinity.
Using the symmetry of the current fluctuation theorem we were able to
simplify the original expressions (\ref{RC2}).
We found a simple structure in terms of $(n-1)/2$ or $(n/2)$
different tensors if $n$ is respectively odd or even.
These tensors characterize the fluctuations of the currents and their
response to affinities up to an order inferior to the response
coefficient. The numerical coefficients pondering the different
tensors turned out to be independent of $n$,
so that the same expressions arise independently of $n$ and of its parity.
Nevertheless, a difference arises between odd and even response coefficients:
for $n$ even there exists a totally symmetric part arising in the
expression of $C^{(n)}$ that does not appear for $n$ odd.

For example, using (\ref{RCeven}) Onsager's coefficients are given by
\begin{equation}
C_{\alpha\beta}^{(1)} = \gamma_1 \ Q_{\alpha\beta,}^{(2,2)}= - 
\frac{1}{2} \ Q_{\alpha\beta,}^{(2,2)} =
L_{\alpha\beta}
\end{equation}
according to Eq. (\ref{YZ}) in terms
of the time correlation
functions of the instantaneous currents \cite{G52,K57} or the
corresponding Helfand moments \cite{H60}.
Onsager's symmetry
\cite{O31} is therefore verified.
Here, the statistical average is carried out with respect to the
state of thermodynamic equilibrium.

The second-order response is expressed as
\begin{eqnarray}
C_{\alpha\beta\gamma}^{(2)} &=& \gamma_1 \ 
Q_{\alpha\{\beta\gamma\}_1}^{(2,3)} = - \frac{1}{2} \
Q_{\alpha\{\beta\gamma\}_1}^{(2,3)}
= - \frac{1}{2} \ \Big[ Q_{\alpha\beta,\gamma}^{(2,3)}  +
Q_{\alpha\gamma,\beta}^{(2,3)} \Big]
\label{RC3}
\end{eqnarray}
which gives the response coefficients $C_{\alpha\beta\gamma}^{(2)}$
in terms of the expressions (\ref{R}):
\begin{equation}
Q_{\alpha\beta,\gamma}^{(2,3)} = - R_{\alpha\beta,\gamma}
\end{equation}

The third-order response coefficients are given by
\begin{eqnarray}
C_{\alpha\beta\gamma\delta}^{(3)} &=& \nonumber \gamma_1 \ 
Q_{\alpha\{\beta\gamma\delta\}_1}^{(2,4)}
+ \gamma_3 \ Q_{\alpha\beta\gamma\delta}^{(4,4)}  \\
&=& - \frac{1}{2} \ \Big[ Q_{\alpha\{\beta\gamma\delta\}_1}^{(2,4)}
-\frac{1}{2} \ Q_{\alpha\beta\gamma\delta}^{(4,4)}  \Big]
\label{RC4}
\end{eqnarray}
where $Q^{(4,4)}$ is the cumulant of order $4$ calculated at
equilibrium by Eqs. (\ref{Ssym.moments}) and $Q^{(2,4)}$ are the
second derivatives of the power spectra with respect to the
affinities at equilibrium by Eq. (\ref{T.correl}):
\begin{eqnarray}
Q_{\alpha\beta\gamma\delta}^{(4,4)} &=& -2 S_{\alpha\beta\gamma\delta} \\
Q_{\alpha\{\beta\gamma\delta\}_1}^{(2,4)} &=& -
T_{\alpha\beta,\gamma\delta}- T_{\alpha\gamma,\beta\delta} -
T_{\alpha\delta,\beta\gamma}
\end{eqnarray}

To illustrate the elimination of the $Q^{(k,n)}$ with $k$ even in a
non trivial case, let us consider the fourth-order response
coefficients. They are given by
\begin{eqnarray}
C_{\alpha\beta\gamma\delta\epsilon}^{(4)} &=&
- \frac{1}{2} \ \sum_{p=1}^{4} \
Q_{\alpha\{\beta\gamma\delta\epsilon\}_p}^{(1+p,5)} \nonumber \\
&=& - \frac{1}{2} \ \Big[
Q_{\alpha\{\beta\gamma\delta\epsilon\}_1}^{(2,5)} +
Q_{\alpha\{\beta\gamma\delta\epsilon\}_2}^{(3,5)}
+
Q_{\alpha\{\beta\gamma\delta\epsilon\}_3}^{(4,5)}  \Big] \nonumber \\
&=& - \frac{1}{2} \ \Big[
Q_{\alpha\{\beta\gamma\delta\epsilon\}_1}^{(2,5)} -\frac{1}{2} \
Q_{\alpha\{\beta\gamma\delta\epsilon\}_3}^{(4,5)}  \Big]
\label{RC5}
\end{eqnarray}
where we used that $Q_{\alpha\beta\gamma\delta\epsilon}=0$ to go from
the first line to the second and the relations (\ref{simpl}) to go
from the second to the third line.

The fourth-order response coefficients can thus be expressed in terms
of the third derivatives of the spectrum
with respect to the affinities and in terms of the first derivative
of the fourth-order correlation functions with respect to the
affinities.
In particular, it presents the expected symmetry
$C_{\alpha\beta\gamma\delta\epsilon}^{(4)} =
C_{\alpha\gamma\beta\delta\epsilon}^{(4)} =
C_{\alpha\epsilon\gamma\delta\beta}^{(4)} =\cdots =
C_{\alpha\beta\gamma\epsilon\delta}^{(4)}$. The expression
(\ref{RC5}) could have been obtained immediately using the general
form (\ref{RCodd}) with coefficients $\gamma_i$ given by
(\ref{gammanum}). One can also note that the coefficients $\gamma_i$
are the same as in expression (\ref{RC3}) as it should.

In the same way, the fifth-order response coefficients is immediately
given by Eq. (\ref{RCeven}) which reads
\begin{eqnarray}
C_{\alpha\beta\gamma\delta\epsilon\mu}^{(5)} = - \frac{1}{2} \ \Big[
Q_{\alpha\{\beta\gamma\delta\epsilon\mu\}_1}^{(2,6)} - \frac{1}{2} \
Q_{\alpha\{\beta\gamma\delta\epsilon\mu\}_3}^{(4,6)} + \
Q_{\alpha\beta\gamma\delta\epsilon\mu}^{(6,6)}\Big] \nonumber
\label{RC6temp}
\end{eqnarray}
as can be verified using using relations (\ref{n-1even}) and
(\ref{simpl}) on the expression (\ref{RC2}) of the tensor.\\

Eventually, we can construct in the same way the higher-order 
relations for the fluctuations.
Indeed the reasoning remain unchanged when considering the 
fluctuations and their responses. Using relations ($\ref{simpl}$) 
recursively yields, for odd $m$,
\begin{equation}
Q_{\alpha_1 \cdots \alpha_m, \beta\cdots\sigma}^{(m,n)} =
\gamma_1 \ Q_{\alpha_1 \cdots 
\alpha_m\{\beta\cdots\sigma\}_1}^{(m+1,n)} + \gamma_3 \
Q_{\alpha_1 \cdots \alpha_m\{\beta\cdots\sigma\}_3}^{(m+3,n)}+\cdots
+ \gamma_{n-2} \ Q_{\alpha_1 \cdots 
\alpha_m\{\beta\cdots\sigma\}_{n-m-2}}^{(n-1,n)}
\qquad (n \; \mbox{odd})
\label{RCodd}
\end{equation}
and \begin{equation}
Q_{\alpha_1 \cdots \alpha_m, \beta\cdots\sigma}^{(m,n)} =
\gamma_1 \ Q_{\alpha_1 \cdots 
\alpha_m\{\beta\cdots\sigma\}_1}^{(m+1,n)} + \gamma_3 \
Q_{\alpha_1 \cdots \alpha_m\{\beta\cdots\sigma\}_3}^{(m+3,n)}+\cdots
+ \gamma_{n-1} \
Q_{\alpha_1 \cdots \alpha_m\beta\cdots\sigma}^{(n,n)}  \qquad (n \; 
\mbox{even})
\label{RCeven}
\end{equation}
The fluctuations are thus expressed in terms of the independent 
tensors with the same ponderation as for the response coefficients.


\section{Conclusions}
\label{Conclusions}

In this paper, we have shown that the
fluctuation theorem for the currents (\ref{CFTrap}) or (\ref{CFT})
implies not only Onsager's reciprocity relations
\cite{O31} along
with the Green-Kubo and Einstein-Helfand formulas
\cite{H60,G52,K57,E05}
for the linear response coefficients,
but also further remarkable relations for the nonlinear response
coefficients at arbitrarily high orders.
These results find their
origin in the validity of the fluctuation theorem for the currents
far from equilibrium
in stochastic rate processes.  The obtained
relations are thus the consequences of the microreversibility.

The
response coefficients are defined by expanding the currents crossing
the nonequilibrium system in powers of the affinities (or
thermodynamic forces).  Therefore, the response coefficients are
defined
with respect to the equilibrium state where the affinities
vanish.  Nevertheless, we can estimate the currents further away from
equilibrium if we use an expansion up to high powers of the
affinities.
This explains that we need a general property valid far
from equilibrium, such as the fluctuation theorem
for the currents,
in order to obtain relations on the nonlinear response coefficients
at arbitrary orders.

Typically, the relations described in the
present paper connect quantities measuring the statistical
correlations among $m$ fluctuating cumulative currents to
corresponding quantities among $m-1$ of them with an extra derivative
with respect to an affinity.  The former characterizes the
fluctuations at $m^{\rm th}$ order and the latter the sensitivity of
the fluctuations at the lower $(m-1)^{\rm th}$ order with respect to
the nonequilibrium constraints.  This is the case for instance for
the equality between Eq. (\ref{S.G}) which measures the sensitivity
of the nonequilibrium correlations among three cumulative currents
under changes of an affinity and Eq. (\ref{Ssym.moments}) which
characterizes the fluctuations between four cumulative currents in
the equilibrium thermodynamic state.  This is the feature which is
found in the relations here described.

In conclusion, the theory
developed in the present paper
provides a general framework to formulate the nonlinear response
theory in nonequilibrium processes.
The results here reported have important applications for the
nonlinear response properties of many nonequilibrium systems such as
the chemical and biochemical reactions \cite{AG04}, the full counting
statistics in mesoscopic conductors \cite{AG06a}, the effusion of
ideal gases \cite{CVK06}, and Van den Broeck's demons \cite{VMK05}.
The present results could be especially important in nonequilibrium
systems at the micro- and nano-scales, where the nonlinear response
properties turn out to be dominant \cite{AG06b}.

\vspace{0.3cm}

{\bf Acknowledgments.}
D.~Andrieux is Research Fellow at the F.N.R.S. Belgium.
This research is financially supported by the ``Communaut\'e fran\c
caise de Belgique''
(contract ``Actions de Recherche Concert\'ees'' No. 04/09-312) and
the National Fund for Scientific Research (F.~N.~R.~S. Belgium,
contract F. R. F. C. No. 2.4577.04).


\end{document}